\documentclass[particles,review,accept,oneauthor,pdflatex,10pt,a4paper]{mdpi}
\def\dac{\displaystyle\frac}
\def\[{\left[}
\def\]{\right]}
\def\({\left(}
\def\){\right)}

\usepackage{amsmath}
\usepackage{amssymb}
\usepackage{graphicx}
\newcommand{\diag}{\mathop{\rm diag}\nolimits}

\usepackage{environ}
\NewEnviron{myequation}{%
\begin{equation}
\scalebox{0.88}{$\BODY$}
\end{equation}}

\firstpage{1}
\articlenumber{x}
\doinum{10.3390/------}
\pubvolume{xx}
\pubyear{2018}
\copyrightyear{2018}
\history{Received: 30 January 2018; Accepted: 28 February 2018; Published: date}

\Title{Realistic Compactification Models in Einstein–Gauss–Bonnet Gravity}

\Author{{Sergey Pavluchenko} \orcidA{}}

\AuthorNames{Sergey Pavluchenko}

\address[1]{%
Programa de P\'os-Gradua\c{c}\~ao em F\'isica, Universidade Federal do Maranh\~ao (UFMA), 65085-580 S\~ao Lu\'is, Maranh\~ao, Brazil; sergey.pavluchenko@gmail.com\\
}

\abstract{We report the results of a study on the dynamical compactification of spatially flat cosmological models in Einstein–Gauss–Bonnet gravity. The~analysis was performed in the arbitrary dimension in order to be more general. We consider both vacuum and $\Lambda$-term cases. Our results suggest that for vacuum case, realistic compactification into the Kasner (power law) regime occurs with any number of dimensions ($D$),
while the compactification into the exponential solution occurs only for $D \geqslant 2$. For the $\Lambda$-term case only compactification into the exponential solution exists, and it only occurs for $D \geqslant 2$
as well. Our results, combined with the bounds on  Gauss--Bonnet coupling and the $\Lambda$-term ($\alpha, \Lambda$, respectively)  from other considerations, allow for the tightening of the existing constraints and forbid $\alpha < 0$.
}


\begin{document}



\section{Introduction}

It is not widely known, but the idea of extra dimensions preceded the concept of general relativity. Indeed, the~first extra-dimensional model was proposed by Nordstr\"om in 1914~\cite{Nord1914}, unifying Nordstr\"om's second
gravity theory~\cite{Nord_2grav} with Maxwell's electromagnetism. With time it became apparent that the true gravity theory was Einstein's, and Kaluza~\cite{KK1} proposed a similar model based on General Relativity (GR):
 in his model five-dimensional Einstein equations could be decomposed into four-dimensional Einstein equations, in addition to Maxwell's electromagnetism. After that, Klein \cite{KK2, KK3} proposed~ a  quantum mechanical interpretation of this
extra dimension, and hence the Kaluza–Klein was formally formulated. Remarkably, their theory unified all
known interactions at that time. With time, more interactions were known and it became clear that to unify them all, more extra dimensions are needed. Nowadays, one of the promising theories for unifying
all interactions is M/string theory.

The gravitational counterpart of M/string theories often has the curvature-squared corrections in the Lagrangian to counter specific ghosts and ``badly-behaving'' kinetic terms. It was demonstrated~\cite{zwiebach}
that the only combination of quadratic terms that leads to a ghost-free nontrivial gravitation interaction is the Gauss--Bonnet (GB) term:

$$
L_{GB} = L_2 = R_{\mu \nu \lambda \rho} R^{\mu \nu \lambda \rho} - 4 R_{\mu \nu} R^{\mu \nu} + R^2.
$$

Zumino~\cite{zumino} extended Zwiebach's result on
higher-than-squared curvature terms, supporting the idea that the low-energy limit of the unified
theory might have a Lagrangian density as a sum of contributions of different powers of curvature. In~this regard, Einstein--Gauss--Bonnet (EGB) gravity could be seen as a subcase of the more general Lovelock gravity~\cite{Lovelock}.

All extra-dimensional theories have one thing in common---the need to explain where additional dimensions are ``hiding'', as we do not sense them, at least with the current level of experiments. The~generally accepted answer is that the additional dimensions are ``compactified'', meaning that they are very small in size. Off course, one would want to make them small naturally, in~the course of evolution and with unsuppressed initial conditions and parameters.

The cosmology with Einstein--Gauss--Bonnet gravity in extra dimensions has been actively studied for some time~\cite{add_4, Deruelle2, Deruelle1}; more recent research focuses on the
power law~\cite{mpla09, prd09, Ivashchuk, prd10, grg10, PT} and\linebreak exponential~\cite{KPT, CPT1, CST2, CPT3, my15, iv16, PT, iv18}~solutions.

In order to find all possible regimes in the Einstein--Gauss--Bonnet cosmology, it is necessary to go beyond an exponential or power law {\it ansatz} and keep the scale factor generic.
As we are particularly interested in models that allow dynamical compactification,  we
consider the metric as being a product of spatially three-dimensional and extra-dimensional parts. In~that case, the three-dimensional part could be seen as ``our Universe'' and we expect this part to expand while the extra-dimensional part should be suppressed in size with respect to the three-dimensional one. In~\cite{CGP1} we demonstrated the existence of a phenomenologically
sensible regime when the curvature of the extra dimensions is negative and the Einstein--Gauss--Bonnet theory does not admit a maximally symmetric solution. In~this case, both the
three-dimensional Hubble parameter and the extra-dimensional scale factor asymptotically tend to the constant values. In~\cite{CGP2} we performed a detailed analysis of the cosmological dynamics in this model
with generic couplings. We recently studied this model in~\cite{CGPT} and demonstrated that, with an additional constraint on couplings, Friedmann-type late-time behavior
could be restored.
Overall, the~consideration of nonzero spatial curvature proved to be worthwhile. For example, for the inflaton field the spatial curvature changes the possibilities for reaching the inflation asymptotes~\cite{infl1, infl2}; in EGB gravity the spatial curvature changes the cosmological regimes~\cite{PT2017}

The current paper is a semi-review: we review the results of~\cite{my16a,my16b, my17a}, but also fix several inaccuracies while providing a new presentation of the results and a discussion of newly found regimes.
Indeed, the~original papers~\cite{my16a,my16b, my17a} have a lot of technical details, which prove the correctness of the results, but, on the other hand, the~physical meaning could be easily lost behind these details. In~the
current manuscript we want to focus on the meaning and not on the technical details. We provide more explanations as to a more elegant way to represent the resulting regimes. In~addition, it appears that one of the power law
regimes in the original papers was mistaken for another; we fix this in the current manuscript (let us note that this mistake does not change the resulting successful compactification regimes, nor does it change the ranges of the parameters where they occur).

The structure of the manuscript is as follows. First, we write down general equations of motion for Einstein--Gauss--Bonnet gravity, and then we rewrite them for our {\it ansatz}. In~the following
section we analyze them for all distinct cases which cover all possible number of extra dimensions for the vacuum case. After that we do the same for the $\Lambda$-term case. After that, we discuss the
regimes and the results obtained, and in addition compare our bounds on $(\alpha, \Lambda)$ with those from other considerations.


\section{Equations of Motion}

Lovelock gravity~\cite{Lovelock} has a structure as follows. Its Lagrangian is constructed from the terms

\begin{equation}
L_n = \frac{1}{2^n}\delta^{i_1 i_2 \dots i_{2n}}_{j_1 j_2 \dots
j_{2n}} R^{j_1 j_2}_{i_1 i_2}
 \dots R^{j_{2n-1} j_{2n}}_{i_{2n-1} i_{2n}}, \label{lov_lagr}
\end{equation}

\noindent where $\delta^{i_1 i_2 \dots i_{2n}}_{j_1 j_2 \dots
j_{2n}}$ is the generalized Kronecker delta of the order $2n$.
One can verify that $L_n$ is Euler-invariant in less than $2n$ spatial dimensions, and so it would not give a nontrivial contribution to the equations of motion. Hence, the
Lagrangian density for any given number of spatial dimensions is sum of all Lovelock invariants (\ref{lov_lagr})  up to  $n=\[\dac{D}{2}\]$ providing nontrivial contributions to the equations of~motion, e.g.,

\begin{equation}
{\cal L}= \sqrt{-g} \sum_n c_n L_n, \label{lagr}
\end{equation}

\noindent where $g$ is the determinant of metric tensor,
$c_n$ are coupling constants of the order of Planck length in $2n$
dimensions, and summation over all $n$ in consideration is assumed. The~most general flat anisotropic metric {\it ansatz} (Bianchi-I-type) has the form

\begin{equation}\label{metric}
g_{\mu\nu} = \diag\{ -1, a_1^2(t), a_2^2(t),\ldots, a_n^2(t)\}.
\end{equation}

As we mentioned earlier, we are interested in the dynamics in quadratic Lovelock (Einstein--Gauss--Bonnet) gravity , so we consider $n$ up to two ($n=0$ is the boundary term while $n=1$ is Einstein--Hilbert and $n=2$ is Gauss--Bonnet gravity).
Substituting metric (\ref{metric}) into the Lagrangian and following the usual procedure gives us the equations of motion:

\begin{equation}
\begin{array}{l}
2 \[ \sum\limits_{j\ne i} (\dot H_j + H_j^2)
+ \sum\limits_{\substack{\{ k > l\} \\ \ne i}} H_k H_l \] + 8\alpha \[ \sum\limits_{j\ne i} (\dot H_j + H_j^2) \sum\limits_{\substack{\{k>l\} \\ \ne \{i, j\}}} H_k H_l +
3 \sum\limits_{\substack{\{ k > l >  \\   m > n\} \ne i}} H_k H_l H_m H_n \] = \Lambda
\end{array} \label{dyn_gen}
\end{equation}

\noindent as the $i$th dynamical equation. The~first Lovelock term, the Einstein--Hilbert contribution, is in the first set of brackets; the~second term, the Gauss--Bonnet term, is in the second set; and $\alpha$
is the coupling constant for the Gauss--Bonnet contribution, while the corresponding constant for the Einstein--Hilbert contribution is put to unity.
Also, since in this section we consider spatially flat cosmological models, scale
factors do not hold much in the physical sense and the equations of motion are rewritten in terms of the Hubble parameters $H_i = \dot a_i(t)/a_i(t)$. Apart from the dynamical equations, we write down the constraint~equation

\begin{equation}
\begin{array}{l}
2 \sum\limits_{i > j} H_i H_j + 24\alpha \sum\limits_{\substack{i > j >\\  k > l}} H_i H_j H_k H_l = \Lambda.
\end{array} \label{con_gen}
\end{equation}

As mentioned in the Introduction,
we want to investigate a particular case with the scale factors split into two parts---the three separate dimensions (a three-dimensional isotropic subspace), which are supposed to represent our world, and~the extra dimensions (the $D$-dimensional isotropic subspace). Hence, we use $H_1 = H_2 = H_3 = H$ and $H_4 = \ldots = H_{D+3} = h$ ($D$ designs the number of extra dimensions), and the
equations take the following forms: the
dynamical equation that corresponds to $H$,
\begin{equation}
\begin{array}{l}
2 \[ 2 \dot H + 3H^2 + D\dot h + \dac{D(D+1)}{2} h^2 + 2DHh\] + 8\alpha \[ 2\dot H \(DHh + \dac{D(D-1)}{2}h^2 \) + \right. \\ \\ \left. + D\dot h \(H^2 + 2(D-1)Hh + \dac{(D-1)(D-2)}{2}h^2 \) +
2DH^3h + \dac{D(5D-3)}{2} H^2h^2 + \right. \\ \\ \left. + D^2(D-1) Hh^3 + \dac{(D+1)D(D-1)(D-2)}{8} h^4 \]
 - \Lambda=0,
\end{array} \label{H_gen}
\end{equation}

\noindent the dynamical equation that corresponds to $h$,
\begin{myequation}
\begin{array}{l}
2 \[ 3 \dot H + 6H^2 + (D-1)\dot h + \dac{D(D-1)}{2} h^2 + 3(D-1)Hh\] + 8\alpha \[ 3\dot H \(H^2 + 2(D-1)Hh + \right. \right. \\ \\ \left. \left. + \dac{(D-1)(D-2)}{2}h^2 \) +  (D-1)\dot h \(3H^2 + 3(D-2)Hh +
\dac{(D-2)(D-3)}{2}h^2 \) + 3H^4 + 9(D-1)H^3h +
\right. \\ \\ \left. +  3(D-1)(2D-3) H^2h^2 +  \dac{3(D-1)^2 (D-2)}{2} Hh^3 + \dac{D(D-1)(D-2)(D-3)}{8} h^4 \] = \Lambda,
\end{array} \label{h_gen}
\end{myequation}

\noindent and the constraint equation,
\begin{myequation}
\begin{array}{l}
2 \[ 3H^2 + 3DHh + \dac{D(D-1)}{2} h^2 \] + 24\alpha \[ DH^3h + \dac{3D(D-1)}{2}H^2h^2 + \dac{D(D-1)(D-2)}{2}Hh^3 + \right. \\ \\ \left. + \dac{D(D-1)(D-2)(D-3)}{24}h^4\] = \Lambda.
\end{array} \label{con2_gen}
\end{myequation}

Looking at (\ref{H_gen})--(\ref{con2_gen}) one can see that the structure of the equations depends on the number of extra dimensions $D$ (terms with $(D-1)$ multiplier nullify in $D=1$, and so on).
In order to cover all cases we need to consider $D=1, 2, 3$ and general $D \geqslant 4$ cases separately. We also consider vacuum and $\Lambda$-term cases separately; for the former we just put $\Lambda=0$ in Equations
(\ref{H_gen})--(\ref{con2_gen}).

\section{The Vacuum Case}

First we consider the  vacuum case. We put $\Lambda=0$ to Equations (\ref{H_gen})--(\ref{con2_gen}) and consider cases with $D=1, 2, 3$ and general $D \geqslant 4$ separately. Since the procedure is generally the same in both vacuum and
$\Lambda$-term cases as well as in all $D$ (with the details differing slightly in various $D$), we present the analysis for the vacuum $D=1$ case in detail and for the rest of the cases we omit these details and provide only the results and their discussions.

\subsection{$D=1$ Case}

In this case the equations of motion take form ($H$-equation, $h$-equation, and constraint, respectively):
\begin{equation}
\begin{array}{l}
4\dot H + 6H^2 + 2\dot h + 2h^2 + 4Hh + 8\alpha \( 2(\dot H + H^2)Hh + (\dot h + h^2)H^2\) = 0,
\end{array} \label{D1_H}
\end{equation}
\begin{equation}
\begin{array}{l}
6\dot H + 12H^2 + 24\alpha (\dot H + H^2)H^2 = 0,
\end{array} \label{D1_h}
\end{equation}
\begin{equation}
\begin{array}{l}
6H^2 + 6Hh + 24\alpha H^3h = 0.
\end{array} \label{D1_con}
\end{equation}

\begin{figure}[H]
\centering
\includegraphics[width=0.8\textwidth, angle=0]{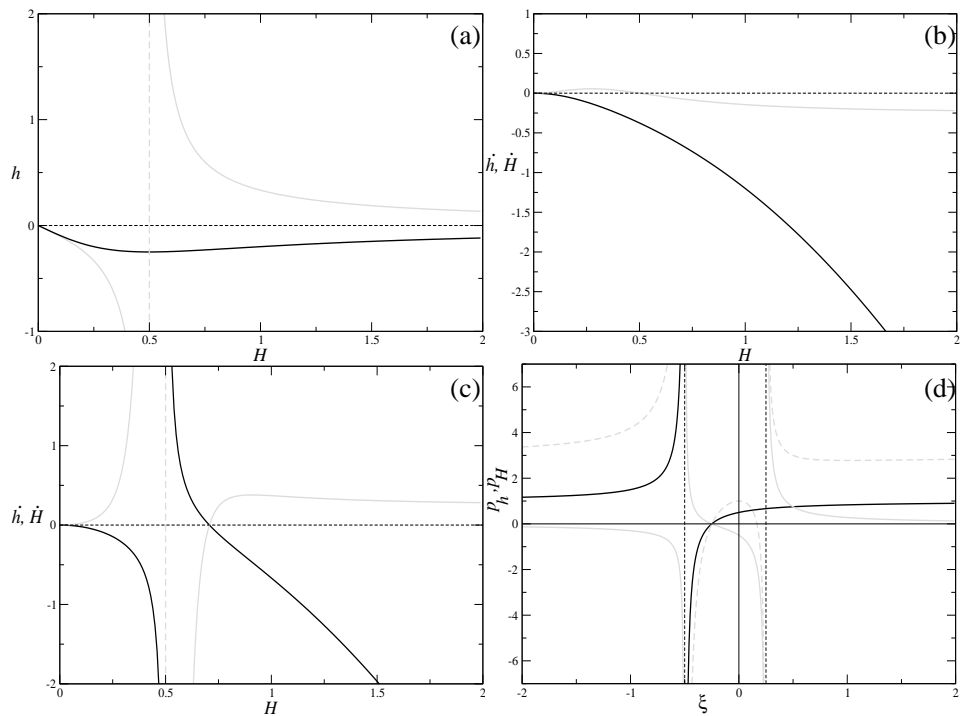}
\caption{Graphs illustrating the dynamics of the $D=1$ vacuum cosmological model. In~panel (\textbf{a}) we present the behavior for $h(H)$ from (\ref{D1_hh}). Black is used for $\alpha > 0$ and gray for $\alpha < 0$. In~panels
(\textbf{b}) and (\textbf{c}) we present $\dot H(H)$ in black and $\dot h(H)$ in gray for $\alpha > 0$ (panel (\textbf{b})) and $\alpha < 0$ (panel (\textbf{c})). Finally, in panel (\textbf{d}) we present the Kasner exponents $p_H$ in black, $p_h$ in
gray, and the expansion rate $(3p_H + p_h)$ in dashed gray; irregularities are denoted as dashed black lines (see the text for more details).}\label{D1v0}
\end{figure}

From (\ref{D1_con}) we can easily see that
\begin{equation}
\begin{array}{l}
h = - \dac{H}{1+4\alpha H^2},
\end{array} \label{D1_hh}
\end{equation}
and so $H$ and $h$ have opposite signs for $\alpha > 0$, but could have same sign in the $\alpha < 0$ case. We presented them in Figure~\ref{D1v0}a---black for $\alpha > 0$ and grey for
$\alpha < 0$. Also, one can resolve Equation (\ref{D1_h}) for the vacuum case with respect to
$\dot H$ to obtain
\begin{equation}
\begin{array}{l}
\dot H = - \dac{2H^2 (1+2\alpha H^2)}{1+4\alpha H^2};
\end{array} \label{D1_dH}
\end{equation}

\noindent and after that with use of (\ref{D1_dH}) one can solve (\ref{D1_H}) to get
\begin{equation}
\begin{array}{l}
\dot h = - \dac{2H^2 (8\alpha^2 H^4 +  2\alpha H^2 - 1)}{(1+4\alpha H^2)(16\alpha^2 H^4 + 8\alpha H^2 + 1)}.
\end{array} \label{D1_dh}
\end{equation}

Now we can plot $\dot H$ and $\dot h$ versus $H$; these values are depicted in Figure~\ref{D1v0}b,c. Panel (b) corresponds to the $\alpha > 0$ case and panel (c) to $\alpha < 0$; particular
curves correspond to $\alpha = \pm 1$.
In these panels we presented $\dot H(H)$ in black and $\dot h(H)$ in gray.
Finally we can perform analysis in terms of the Kasner exponents $p_i = - H_i^2/\dot H_i$; it is convenient if we want to find power law ($a(t) \propto t^{p}$) asymptotes. With $\dot H_i$ taken from Equations (\ref{D1_dH})--(\ref{D1_dh}),
the Kasner exponents could be easily written down as

\begin{equation}
\begin{array}{l}
p_H = \dac{1}{2}\times\dac{4\xi + 1}{2\xi + 1},~p_h = \dac{1}{2}\times\dac{4\xi + 1}{8\xi^2 + 2\xi - 1}~\mbox{with}~\xi = \alpha H^2.
\end{array} \label{D1_pHph}
\end{equation}

We plot the resulting curves in Figure~\ref{D1v0}d, with $p_H$ represented with a black line, $p_h$ in gray and their sum $\sum p_i = 3p_H + p_h$ in dashed gray

Now with all the preliminaries done we can describe regimes. For $\alpha > 0$ (black line in Figure~\ref{D1v0}a, whole Figure~\ref{D1v0}b, and $\xi > 0$ part of the Figure~\ref{D1v0}d) we can conclude that there is transition from the
$P_{(1, 0)}$ power law regime with $p_H = 1$ and $p_h = 0$, to the low-energy (standard) Kasner regime $K_1$ with $\sum p = 1$; we shall discuss the notations a bit later.

For $\alpha < 0$ (gray line in Figure~\ref{D1v0}a, whole Figure~\ref{D1v0}c and $\xi < 0$ part of the Figure~\ref{D1v0}d) we have three regimes. The first is the high-energy $P_{(1, 0)}$ regime transit into the isotropic exponential
solution~{(see~\cite{iv10}  for $\Lambda = 0$ and \cite{CPT1} for $\Lambda \neq 0$) }.
$E_{iso}$ at $H^2 = H_1^2 = -1/(2\alpha)$. The~second (by decreasing absolute value of $H$) is the transition from the nonstandard singularity at $H^2 = H_2^2 = -1/(4\alpha)$ to the same isotropic exponential solution as above.
We comment more on the nonstandard singularities in the Discussion section; here we just mention the situation when $\dot H$ or $\dot h$ diverges at the final $H$ and/or $h$, creating physical singularity at
final and nonzero $H$ and/or $h$; we denote these values as $nS$.
The third regime is the transition from the same nonstandard singularity to low-energy Kasner $K_1$.

In order to describe all possible regimes for this case---for $\alpha > 0$ regardless of the initial $H_0$---we always have the same transition $P_{(1, 0)}\to K_1$,so that for all initial $H_0 > 0$, the past asymptote would be
$P_{(1, 0)}$, while the future asymptote is $K_1$. For $\alpha < 0$ the situation is different---there are three regimes separated by the isotropic exponential solution at $H_1$ and nonstandard singularity at $H_2$. Hence,
for $H > H_1$ we have a $P_{(1, 0)}\to E_{iso}$ transition (for all initial $H_0 > H_1$ the past asymptote would be $P_{(1, 0)}$ while the future asymptote is $E_{iso}$), for $H_1 > H > H_2$ the regime is $nS \to E_{iso}$, and
finally
for $H_2 > H > 0$ the regime is $nS \to K_1$.

\begin{figure}
\centering
\includegraphics[width=0.9\textwidth, angle=0]{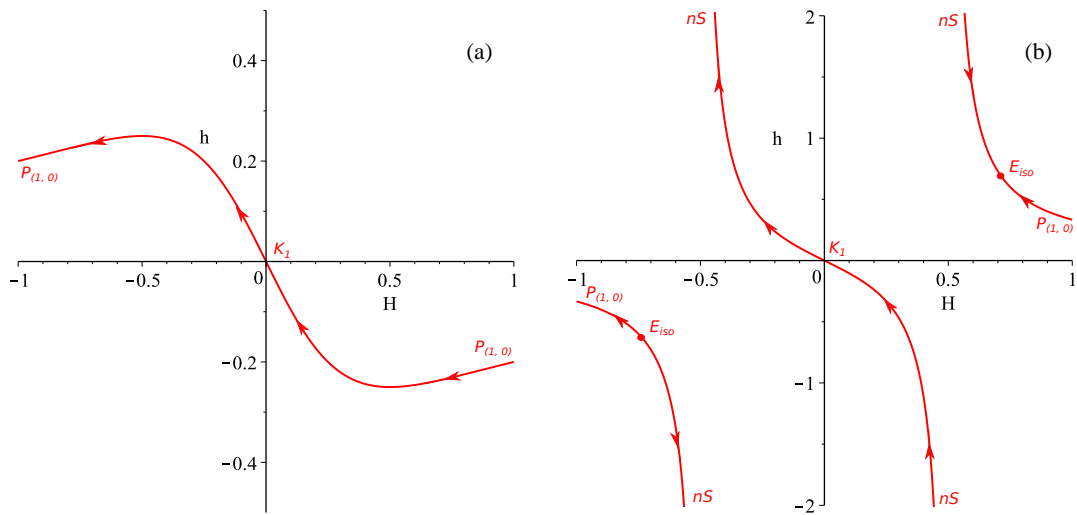}
\caption{ ({\bf{a}})--({\bf{b}}) The regimes shown in the resulting $H(h)$ graph for the $D=1$ vacuum case (see the text for more~details).}\label{D1v}
\end{figure}

One can see that all possible combinations of $\alpha$ and $H_0$ (these are the only independent parameters and are the initial conditions for a well-posed Cauchy problem) fall within the described above four cases, and so give a complete description of the dynamics of this case. For illustrative purposes we combine them in Figure~\ref{D1v}---$\alpha > 0$ in panel (a)  and $\alpha < 0$ in panel (b). The~direction of the evolution is designed with arrows.
In all future cases we shall give the resulting regimes
exactly on the $H(h)$ (or~$h(H)$~for high-$D$~cases) graphs.

Now is an appropriate time to introduce and discuss the designations for the regimes we use through the paper. We design exponential solutions with the letter $E$, with the subindex distinguishing between isotropic ($E_{iso}$)
and anisotropic ($E_{3+D}$) solutions. The~former  corresponds to the case where all spatial dimensions expand isotropically; of course this is not what we observe (we observe only three spatial dimensions to expand)
so we usually treat $E_{iso}$ as a non-singular yet non-realistic regime. In contrast, $E_{3+D}$ is an anisotropic exponential solution with different expansion rates in three and $D$ dimensions (the actual notation
is different in each $D$, for example, $E_{3+2}$ in $D=2$ and so on); but to be treated as realistic we require both $H > 0$ (the expansion of three-dimensional subspace) and $h \leqslant 0$ (the extra dimensions that are either constant in size
or contract). We added the latter condition to ensure that three dimensions are much larger than the extra dimensions. Of course, this situation is not fulfilled in all $E_{3+D}$, so we comment about it in each particular case.

The power law regimes are a bit tricky. In~the original papers~\cite{my16a, my16b, my17a} we used only Kasner regimes, but more in-depth investigation revealed that it is not the case and here we provide an updated
classification. Hence, there are two classes of power law solution; the first of them is the Kasner solution. As derived in~\cite{prd09}, in~the Lovelock gravity of the order $n$, the~Kasner solution is defined as
$\sum p_i = (2n-1)$ and $\sum p_1 ... p_{2n} = 0$ (one can check that the standard Kasner solution with $\sum p_1 = \sum p_i^2 = 1$ also follows this rule). With that at hand, Gauss--Bonnet Kasner solution should have
$\sum p = 3$. However, the solution with $p_H = 1$ and $p_h = 0$ also gives $\sum p = 3$ but it is not a Kasner solution (instead it could be seen as generalization of the Taub solution~\cite{Taub}). This misled us in earlier
studies~\cite{my16a, my16b, my17a} so we have fixed it in current paper. Hence, this is the second class of the power law solutions, which we denote as $P$ with a subindex distinguishing between $p_H = 1$, $p_h = 0$ ($P_{(1, 0)}$) and
$p_H = 0$, $p_h = 1$ ($P_{(0, 1)}$). We discuss the possible connection between this and the Taub solution in the Discussion section.

Finally we denote the nonstandard singularity as $nS$---it is a physical singularity which appears in non-linear gravity theories and is due to this non-linearity. We describe and discuss it in the Discussion section; for
now we just mention that it is a singular  and non-viable regime.

To conclude, in~$D=1$ vacuum case there are total four regimes but only two of them are nonsingular---$P_{(1, 0)} \to K_1$ for $\alpha > 0$ and $P_{(1, 0)} \to E_{iso}$ for $\alpha < 0$. Of these two only one
could be called viable---$P_{(1, 0)} \to K_1$ for $\alpha > 0$---since the other one supposes isotropisation of the entire space and this is not what we observe. Still, the viability of $P_{(1, 0)}$ is a big question and we discuss it in the Discussion~section.

\subsection{$D=2$ Case}

The procedure for $D=2$ is generally the same as for $D=1$; the difference is in the fact that in the $D=2$ case the constraint is quadratic with respect to $h$ and so instead of a single $H(h)$ curve we have two branches. For both of them we repeat the procedure and the results are presented in Figure~\ref{vac}a,b; different colors (red and blue) correspond to different branches.
In Figure~\ref{vac}a we present the $\alpha > 0$ case, while in Figure~\ref{vac}b $\alpha < 0$.
All the regimes are presented there but let us focus only on the realistic ones: for $\alpha > 0$ (Figure~\ref{vac}a),  $K_3 \to E_{3+2}$ and $P_{(1, 0)}\to K_1$. We see that in $D=2$ we finally have the regular Gauss--Bonnet
Kasner
solution. The~exponential solution $E_{3+2}$, being located in the fourth quadrant of Figure~\ref{vac}a, has $H > 0$ and $h < 0$ and thus is a viable regime. The~same is true for $K_1$ as well---for it we also
have $H > 0$ and $h < 0$. For $\alpha < 0$ (Figure~\ref{vac}b) the only realistic regime is $K_3 \to K_1$.

To conclude the vacuum $D=2$ case, we finally have a regular Gauss--Bonnet Kasner solution $K_3$ and a viable anisotropic exponential solution $E_{3+2}$; the list of realistic regimes includes $K_3 \to E_{3+2}$ and $P_{(1, 0)}\to K_1$
for
$\alpha > 0$ and $K_3 \to K_1$ for $\alpha < 0$. Let us also note that $P_{(1, 0)}\to K_1$ in the $D=1$ vacuum case also occurs for $\alpha < 0$.

\subsection{$D=3$ Case}

\begin{figure}[H]
\centering
\includegraphics[width=0.85\textwidth, angle=0]{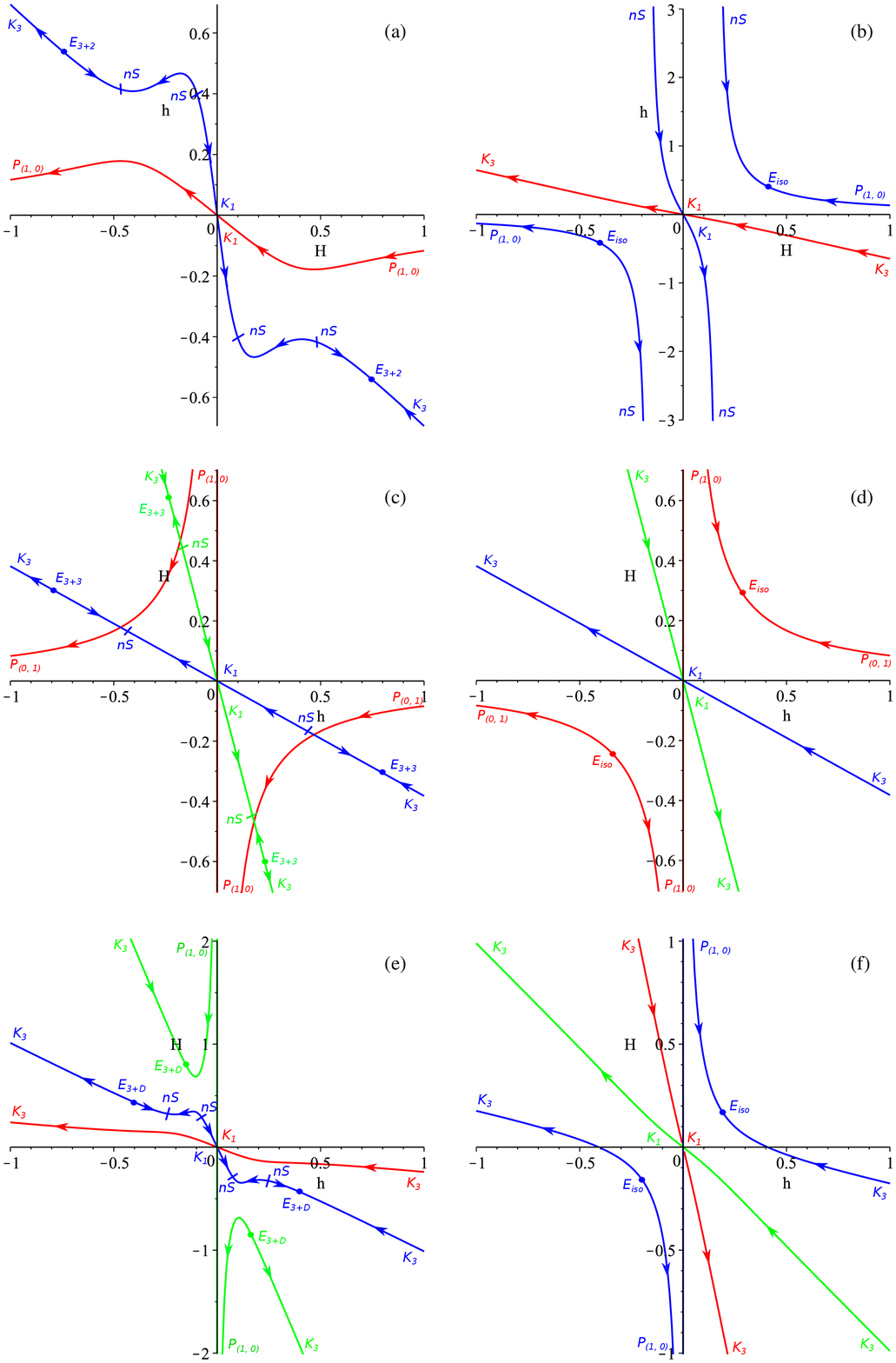}
\caption{The regimes shown in the resulting $H(h)$ graph for the $D=2$ vacuum case (panels (\textbf{a}) and (\textbf{b})), the $D=3$ vacuum case (panels (\textbf{c}) and (\textbf{d})), and the general $D \geqslant 4$ case (panels (\textbf{e}) and (\textbf{f}))
(see the text for more details).}\label{vac}
\end{figure}

This case is also similar to the previous ones and its complexity has also grown a bit---now the constraint equation is cubic with respect to both $H$ and $h$. Considering that in the general $D \geqslant 4$ case it is
quartic with respect to $h$ and cubic with respect to $H$, we switch the variable we solve the constraint and for $D \geqslant 3$ solve it with respect to $H$. Starting with this case we have three branches for
$h(H)$ curves; the axes in Figure~\ref{vac} are also swapped.

The $D=3$ case has an unique feature. Since the number of extra dimensions is three---the same as the number of our spatial dimensions---it is irrelevant which three are expanding and which three are contracting, so we have ``double'' the number of realistic regimes. Indeed, as we can see from  Figure~\ref{vac}c ($D=3$ case with $\alpha > 0$), there are two distinct $K_3 \to E_{3+3}$ regimes---on the blue branch in the fourth quadrant
and on the green branch in the second quadrant. One can see that these are different $E_{3+3}$ solutions and they originate from different $K_3$. The~same is true for $K_3 \to K_1$ regimes, which can be found in
Figure~\ref{vac}d ($D=3$ case with $\alpha < 0$) on the blue branch in the fourth quadrant and on the green branch in the second quadrant. One can also check that they are leading to different $K_1$ and originate from
different $K_3$.

It is interesting to see what becomes of the regime originating from $P_{(1, 0)}$. In $D=1, 2$,  $P_{(1, 0)}\to K_1$, but in $D=3$ it becomes $P_{(1, 0)}\to P_{(0, 1)}$. Both asymptotes are power law and
Taub-like, but the subspaces interchange static and expanding parts---initially one of the subspaces almost is static and another is expanding. Then, the expanding starts to slow down while another starts to expand and
finally the initial expansion stops and the initial static expands.

In the vacuum $D=3$ case the situation is similar to the $D=2$ case---we have a realistic  $K_3 \to E_{3+3}$ regime (actually, two) and $P_{(1, 0)}\to P_{(0, 1)}$
 for $\alpha > 0$ (similar to the $D=2$ case), and a realistic $K_3 \to K_1$ regime
for $\alpha < 0$.
\subsection{General $D \geqslant 4$ Case}

The final case is the general case of $D \geqslant 4$. Despite the fact that actual curves are a bit different for each particular $D$, the~shape, derivatives, and so on are quite similar, so we consider the general $D \geqslant 4$ case with
$D=4$ as an example, which can be seen in Figure~\ref{vac}e,f. Similar to the previous cases, the left panel (Figure~\ref{vac}e) is for $\alpha > 0$ while the right panel (Figure~\ref{vac}f) is for $\alpha < 0$; different colors
correspond to different branches in the sense of the solutions in the constraint Equation (\ref{con2_gen}) with respect to $H$.

As we defined the realistic regime as one with $H > 0$ and $h \leqslant 0$, the~regimes vital for us should be located in the second quadrant. One can see that in the fourth quadrant we have the $K_3 \to E_{3+D}$ regime which
sounds good, but the final anisotropic exponential solution has $H < 0$ and $h > 0$. So, despite the fact that the regime is nonsingular, the~final state is unrealistic. In the second quadrant of Figure~\ref{vac}e
we can see two realistic regimes---$K_3 \to E_{3+D}$ and $P_{(1, 0)} \to E_{3+D}$---these regimes correspond to $\alpha > 0$. For $\alpha < 0$, which could be found in Figure~\ref{vac}f, the~only realistic regime is
$K_3 \to K_1$ from the second quadrant.

To conclude, the~general $D \geqslant 4$ case is similar to the $D=3$ -- for $\alpha > 0$ it has $K_3 \to E_{3+D}$ and $P_{(1, 0)} \to E_{3+D}$ regimes (the latter is absent in $D=3$) while for $\alpha < 0$ it has
$K_3 \to K_1$.

\subsection{Conclusions on the Vacuum Case}

Overall, the~vacuum case demonstrates some consistency over variation of $D$. In~particular, in~all $D$ for $\alpha > 0$ there is a regime which originates from $P_{(1, 0)}$,  but the past asymptote for this regime differs with $D$.
In $D=1$ and $D=2$ it is $P_{(1, 0)} \to K_1$, in~$D=3$ it is $P_{(1, 0)}\to P_{(0, 1)}$ and in the general $D \geqslant 4$ case it is $P_{(1, 0)} \to E_{3+D}$. Apart from these regimes, there are two more which are
presented in $D\geqslant 2$: $K_3 \to E_{3+D}$ for $\alpha > 0$ and $K_3 \to K_1$ for $\alpha < 0$.

The details of the analysis, such as the exact locations of all exponential solutions and all nonstandard singularities, can be found in~\cite{my16a}, with the corresponding remark on the different treatment of the $P_{(1, 0)}$ regime kept in
mind (in~\cite{my16a} we mistakenly treated it as $K_3$, as we discussed in $D=1$ subsection).

\section{The $\Lambda$-Term Case}

Now, we consider the $\Lambda$-term case. The~procedure is exactly the same as in the previous section, but now we have two parameters of the theory (Gauss--Bonnet coupling $\alpha$ and the $\Lambda$-term). Hence, the classification
becomes more complicated and so there are more cases than in the vacuum case.

\subsection{The $D=1$ Case}

\begin{figure}
\centering
\includegraphics[width=1\textwidth, angle=0]{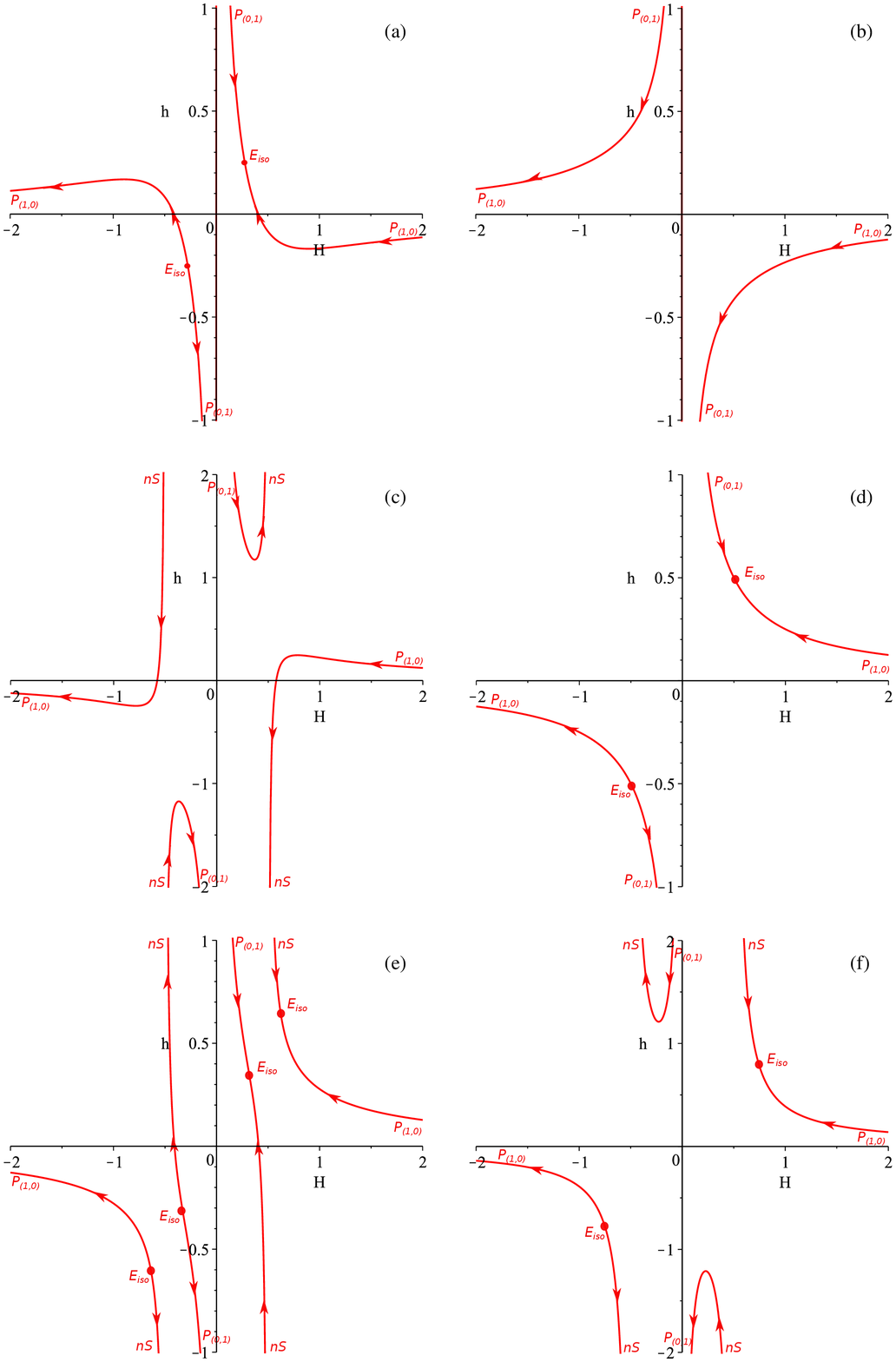}
\caption{({\bf{a}})--({\bf{f}}) The regimes shown in the resulting $h(H)$ graph for the $D=1$ $\Lambda$-term case
(see the text for more~details).}\label{D1L}
\end{figure}

This is the simplest case and, similar to the $D=1$ vacuum case, we have only one branch. The~results are presented in Figure~\ref{D1L} and the panel layout is as follows: in panel (a) there is the
$(\alpha > 0, \Lambda > 0)$ case; in panel (b) there is the $(\alpha > 0, \Lambda < 0)$ case; in panel (c)  $(\alpha < 0, \Lambda > 0. \alpha\Lambda < -3/2)$; in panel (d) $(\alpha < 0, \Lambda > 0. \alpha\Lambda = -3/2)$, in
panel (e) $(\alpha < 0, \Lambda > 0, \alpha\Lambda > -3/2)$ and finally in panel (f) we have presented the $(\alpha < 0, \Lambda < 0)$ case.

Analyzing Figure~\ref{D1L} one can see that there is only one regime which could be called viable--- $P_{(1, 0)}\to P_{(0, 1)}$---which is present in the fourth quadrant ($H > 0, h < 0$) of Figure~\ref{D1L}b (so it is entirely on the $\alpha > 0, \Lambda < 0$ subplane).
 All other
regimes either are located in the ``wrong'' quadrant or have a non-viable past or future asymptote.

Overall, in the $D=1$ $\Lambda$-term case we report one  possible viable regime---$P_{(1, 0)}\to P_{(0, 1)}$---which is the same as in the $D=3$ vacuum case with the difference that now the number of extra dimensions is one.
Later in the Discussion section we discuss it more.
On comparison with the $D=1$ vacuum case one can note that the dynamics are much richer but the number of realistic regimes is the same---only one.

\subsection{The $D=2$ Case}

\begin{figure}
\centering
\includegraphics[width=0.85\textwidth, angle=0]{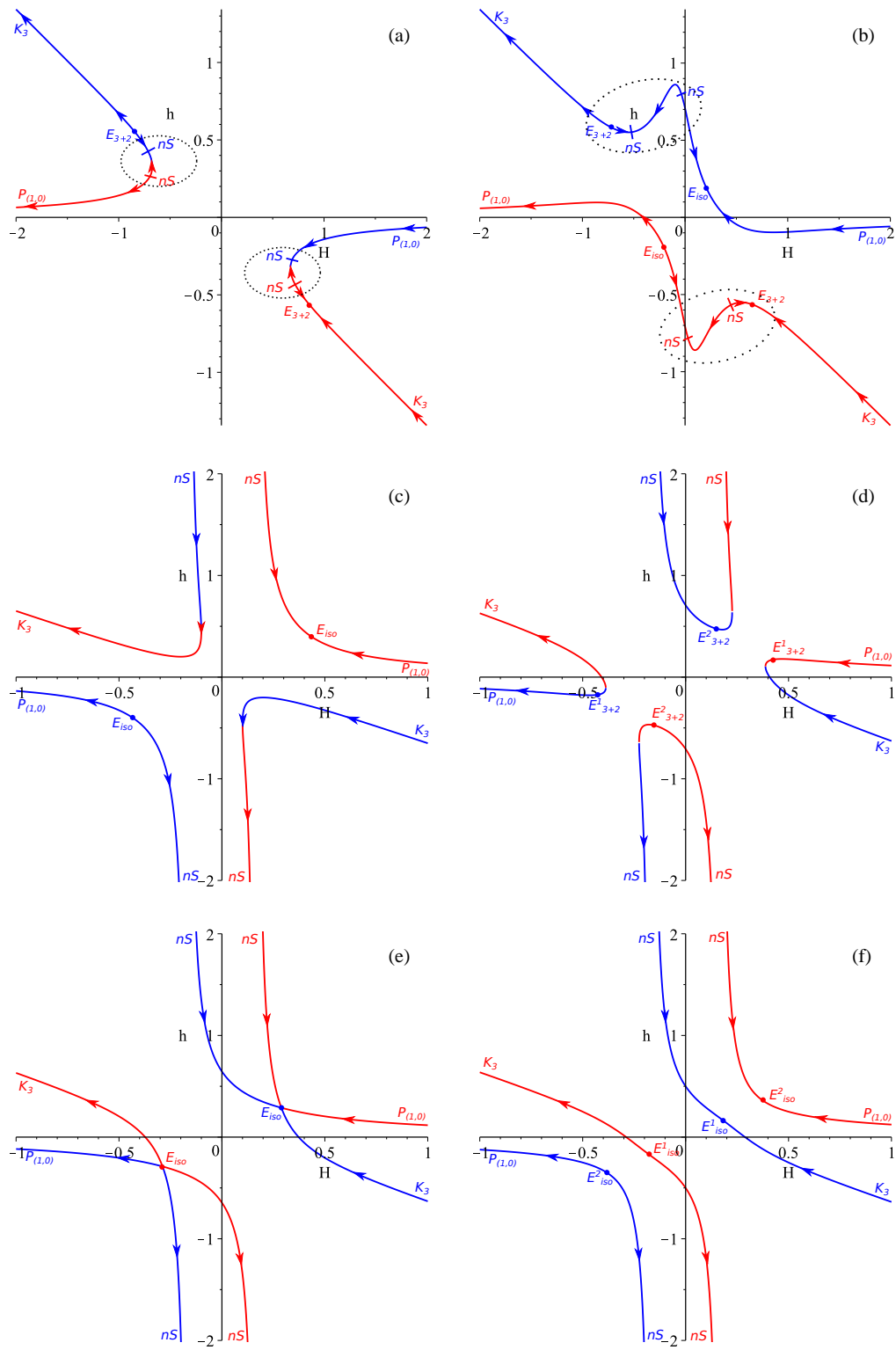}
\caption{({\bf{a}})--({\bf{f}}) The regimes shown in the resulting $h(H)$ graph for the $D=2$ $\Lambda$-term case
(see the text for more~details).}\label{D2L}
\end{figure}

In the $D=2$ $\Lambda$-term case, as in the $D=2$ vacuum case, there are two branches and they are represented as different colors in Figure~\ref{D2L}. The~panel layout is as follows: in panel (a) we presented the
$(\alpha > 0, \Lambda < 0)$ case, in panel (b) the $(\alpha > 0, \Lambda > 0)$ case is shown, in panel (c) the $(\alpha < 0, \Lambda < 0)$ case is presented, and the remaining three panels correspond to the  $(\alpha < 0, \Lambda > 0)$ cases:
$\alpha\Lambda < -5/6$ in panel (d), $\alpha\Lambda = -5/6$ in panel (e) and $\alpha\Lambda > -5/6$ in panel (f).

The first two panels require additional explanations: in Figure~\ref{D2L}a we presented $(\alpha > 0, \Lambda < 0)$ case, but there are changes which separate cases with $\alpha\Lambda \geqslant -1/6$ and
$\alpha\Lambda < -1/6$. As they do not change the shape of the $h(H)$ curves, we decided to keep both of them on the same graph. The~difference is in addition of two nonstandard singularities for $\alpha\Lambda \geqslant -1/6$
---in Figure~\ref{D2L}a they are encircled by a dashed line. For $\alpha\Lambda < -1/6$ they are absent, and so the regimes for the fourth quadrant ($H > 0, h < 0$) are: $P_{(1, 0)} \to E_{3+2} \leftarrow K_3$. In contrast, for $\alpha\Lambda \geqslant -1/6$ they are present and so the regimes are $P_{(1, 0)} \to ns \leftarrow ns \to E_{3+2} \leftarrow K_3$. This way, the~regime $P_{(1, 0)} \to E_{3+2}$, which is realistic, can
exist only for $\alpha\Lambda < -1/6$.

The second panel which requires additional explanations is Figure~\ref{D2L}b. This case corresponds to $(\alpha > 0, \Lambda > 0)$ and there is a fine structure of the anisotropic exponential solutions for\linebreak
$15/32 \geqslant \alpha\Lambda \geqslant 1/2$ (see~\cite{my16b} for details). The~particular example of the situation in Figure~\ref{D2L}b corresponds to $0 < \alpha\Lambda < 15/32$ and for other intervals and exact
values within the $15/32 \geqslant \alpha\Lambda \geqslant 1/2$ range the abundance and locations of the anisotropic exponential solutions and nonstandard singularities are different (see~\cite{my16b} or Discussions
in~\cite{my17a} for details). All of them are ``internal'' in the sense that they cannot be reached from $K_3$ and they do not introduce new realistic regimes. The~regime $K_3 \to E_{3+2}$,
which is realistic, exists for $\alpha\Lambda \leqslant 1/2$; for $\alpha\Lambda > 1/2$ it is replaced with $K_3 \to nS$.

Of the remaining panels we want to briefly comment on Figure~\ref{D2L}d, which corresponds to\linebreak $(\alpha < 0, \Lambda > 0)$, $\alpha\Lambda < -5/6$. The anisotropic exponential solution $E^2_{3+2}$ is located in the
first and third quadrants, but it is so only for $-5/6 > \alpha\Lambda > -3/2$. With the decrease of $\alpha\Lambda$, the location of the solution ``moves'' towards the $H=0$ axis and for $\alpha\Lambda = -3/2$ it is located exactly
at $H=0$. For $\alpha\Lambda < -3/2$ it ``moves'' even further to $H < 0$ on the upper branch and $H > 0$ on the bottom. However, since the past asymptotes for the upper branch are both nonstandard singularities, the~regime
cannot be called viable and we discard it.

The remaining panels do not have realistic regimes so we have only described two in panels (a) and (b): $P_{(1, 0)} \to E_{3+2}$ which exists for $(\alpha > 0, \Lambda < 0)$, $\alpha\Lambda < -1/6$ and
$K_3 \to E_{3+2}$ which exists for $\alpha > 0$, $\alpha\Lambda \leqslant 1/2$ (including the entire $\Lambda < 0$ domain).

Compared with $D=2$ vacuum case, we also have one regime with an anisotropic exponential solution, but lacks power law solutions. In the Discussion section we explain why this is so.

\subsection{The $D=3$ Case}

\begin{figure}
\centering
\includegraphics[width=1.0\textwidth, angle=0]{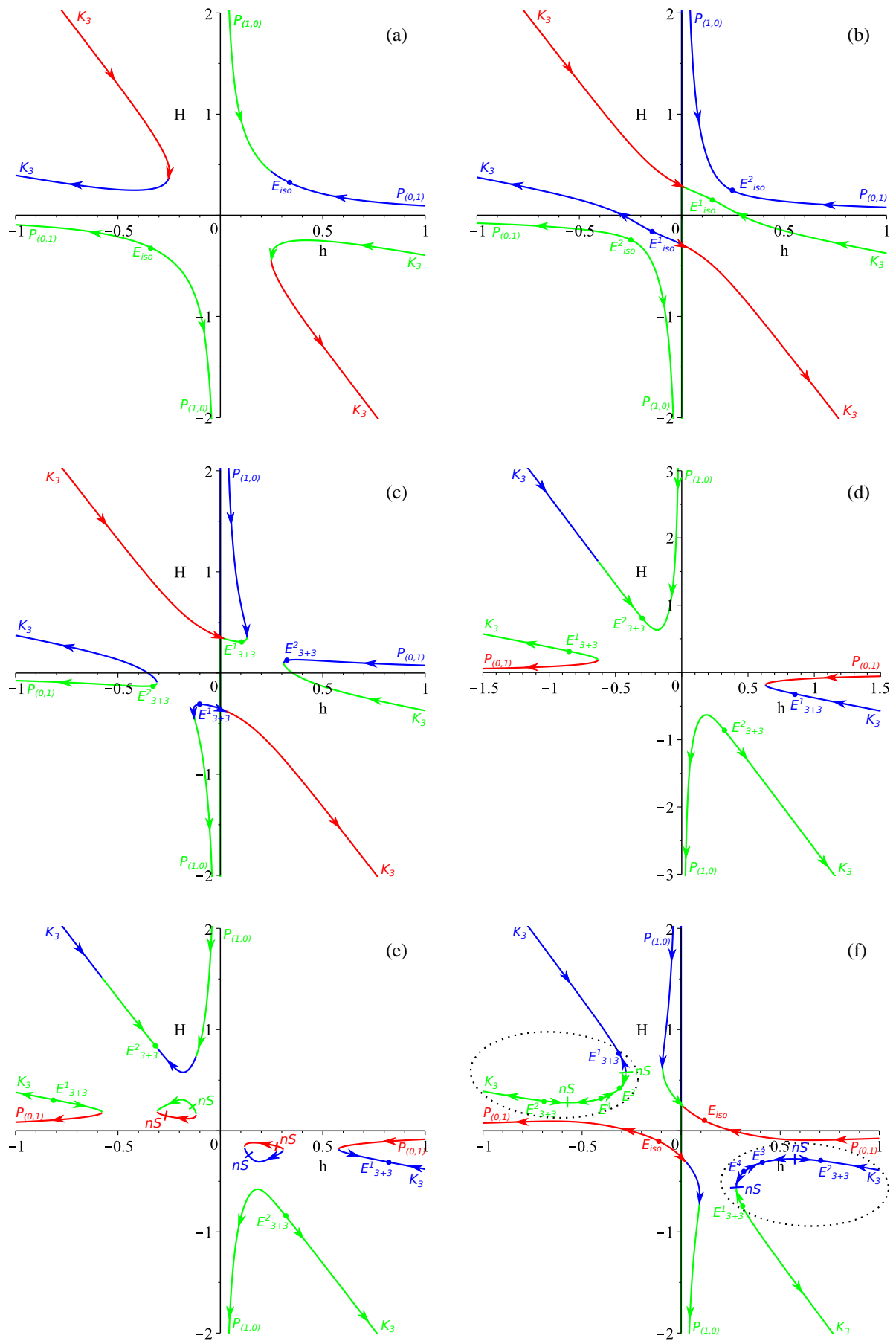}
\caption{({\bf{a}})--({\bf{f}}) The regimes shown in the resulting $H(h)$ graph for the $D=3$ $\Lambda$-term case
(see the text for more~details).}\label{D3L}
\end{figure}

Similar to $D=3$ vacuum case, we have three branches now. Also, as in the $D=3$ vacuum case, both subspaces are three-dimensional, which simplifies the analysis. The~results are presented in Figure~\ref{D3L} and
the panel layout is as follows: in panel (a) we present the
$(\alpha < 0, \Lambda < 0)$ case; in panel (b) the $(\alpha < 0, \Lambda > 0)$, $\alpha\Lambda > -5/8$ case is shown; in panel  (c) we have the $(\alpha < 0, \Lambda > 0)$, $\alpha\Lambda < -5/8$ case; in panel (d), $(\alpha > 0, \Lambda < 0)$,
 $\alpha\Lambda \leqslant -1/8$; in panel (e) $(\alpha > 0, \Lambda < 0)$, $\alpha\Lambda > -1/8$; and finally in panel (f) we have the case of $(\alpha > 0, \Lambda > 0)$. We have not placed here the exact $(\alpha < 0, \Lambda > 0)$,
  $\alpha\Lambda = -5/8$ case, but one can restore it with good precision from Figure~\ref{D3L}b by making two isotropic exponential solutions coincide. As in the case of $(\alpha < 0, \Lambda > 0)$, $\alpha\Lambda > -5/8$, the
  exact $\alpha\Lambda = -5/8$ case does not have any realistic regimes.

We must comment on the $(\alpha > 0, \Lambda > 0)$ case presented in Figure~\ref{D3L}f. Similar to the $D=2$ $\Lambda$-term case, described above, there is a fine structure of the anisotropic exponential solutions, which
can be found in~\cite{my17a}. However, similar to the mentioned case, the~fine structure is ``internal'' and so changes within do not alter realistic regimes. As in the $D=2$ $\Lambda$-term case, there is also the boundary
$\alpha\Lambda \leqslant -1/2$ where the realistic regime $K_3 \to E_{3+3}$ exists.

Before looking for the remaining realistic regimes, let us note that at $D \geqslant 3$, similar to the vacuum case, we solve constraints with respect to $H$ and so swap the axis on the figures; now the realistic regimes
($H > 0, h < 0$) lie in the second quadrant. However, this is also $D=3$ so that only for this case we also can consider ($H < 0, h > 0$)---the fourth quadrant.
There we can find only a few of the regimes. The interesting regime $K_3 \to K_3$ for $(\alpha < 0, \Lambda < 0)$ is presented in Figure~\ref{D3L}a. This is discussed in the Discussion section. In~Figure~\ref{D3L}c---the $(\alpha < 0, \Lambda > 0)$, $\alpha\Lambda < -5/8$ case---the situation is similar to what we described for Figure~\ref{D2L}d. The location of $E_{3+3}^1$ is ``moving'' towards $h=0$ as $\alpha\Lambda$ decreases
and at $\alpha\Lambda = -3/2$ it hits zero (so at $\alpha\Lambda = -3/2$ anisotropic exponential solution $E_{3+3}^1$ has $h=0$). A further decrease of $\alpha\Lambda$ ``moves'' $E_{3+3}^1$ into the second
quadrant for the upper branch and to the fourth quadrant for the lower branch. This way for $\alpha\Lambda \leqslant -3/2$ we have the two realistic regimes $K_3 \to E_{3+3}^1$ and $P_{(1, 0)} \to E_{3+3}^1$.

The other realistic regimes can be found in Figure~\ref{D3L}d ($(\alpha > 0, \Lambda < 0)$, $\alpha\Lambda \leqslant -1/8$)---in the second quadrant they are $K_3 \to E_{3+3}^2$ and $P_{(1, 0)} \to E_{3+3}^2$ while in
the fourth they are $K_3 \to E_{3+3}^1$ and $P_{(0, 1)} \to E_{3+3}^1$. Let us note that $E_{3+3}^1$ and $E_{3+3}^2$ are different exponential solutions, but $P_{(1, 0)}$ and $P_{(0, 1)}$ could be seen as the same. Considering
further Figure~\ref{D3L}e ($(\alpha > 0, \Lambda < 0)$, $\alpha\Lambda > -1/8$) one can see that the realistic regimes are exactly the same.

Finally, in~the $(\alpha > 0, \Lambda > 0)$ case, presented in Figure~\ref{D3L}f, similarly to the previous case we have fine structure of the ``internal'' anisotropic exponential solutions, which leaves the realistic
regime $K_3 \to E_{3+3}$ unaffected. This regime exists for $\alpha\Lambda \leqslant 1/2$---the same value as for the $D=2$ case.

To conclude, there are three potentially realistic regimes in the $D=3$ $\Lambda$-term case: $K_3 \to K_3$ for $(\alpha < 0, \Lambda < 0)$; $P_{(1, 0)} \to E_{3+3}$ for $(\alpha > 0, \Lambda < 0)$ and
$(\alpha < 0, \Lambda > 0, \alpha\Lambda \leqslant -3/2)$; and~$K_3 \to E_{3+3}$ for $(\alpha < 0, \Lambda > 0, \alpha\Lambda \leqslant -3/2)$ and $(\alpha > 0, \alpha\Lambda \leqslant 1/2)$ (including the entire $\Lambda < 0$
domain).

Compared with $D=3$ vacuum case, we have richer dynamics in the $\Lambda$-term case.

\subsection{General $D \geqslant 4$ Case}

\begin{figure}
\centering
\includegraphics[width=0.8\textwidth, angle=0]{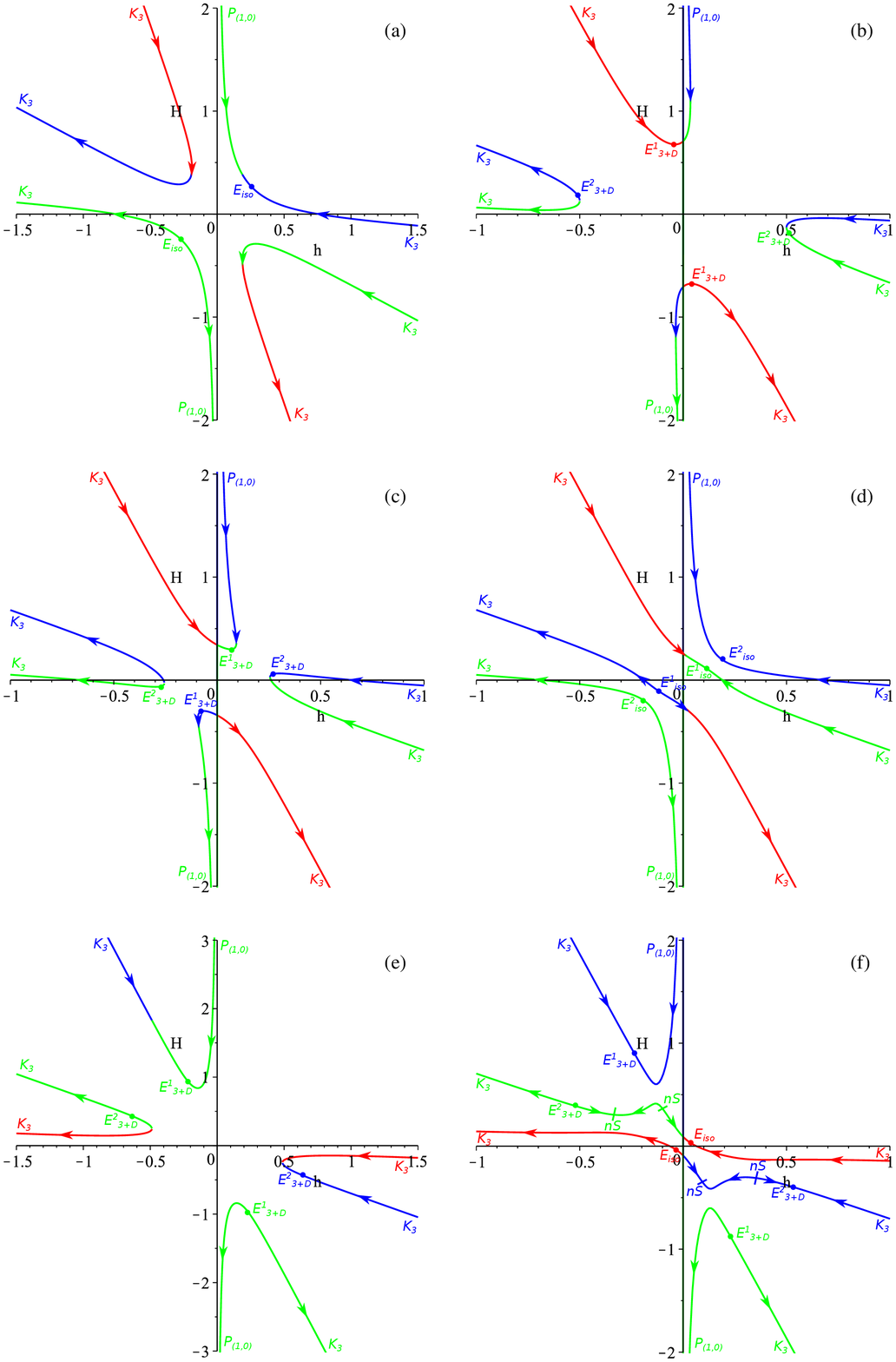}
\caption{({\bf{a}})--({\bf{f}}) The regimes shown in the resulting $H(h)$ graph for the $D\geqslant 4$ $\Lambda$-term case
(see the text for more~details).}\label{D4L}
\end{figure}

\begin{figure}
\centering
\includegraphics[width=1.0\textwidth, angle=0]{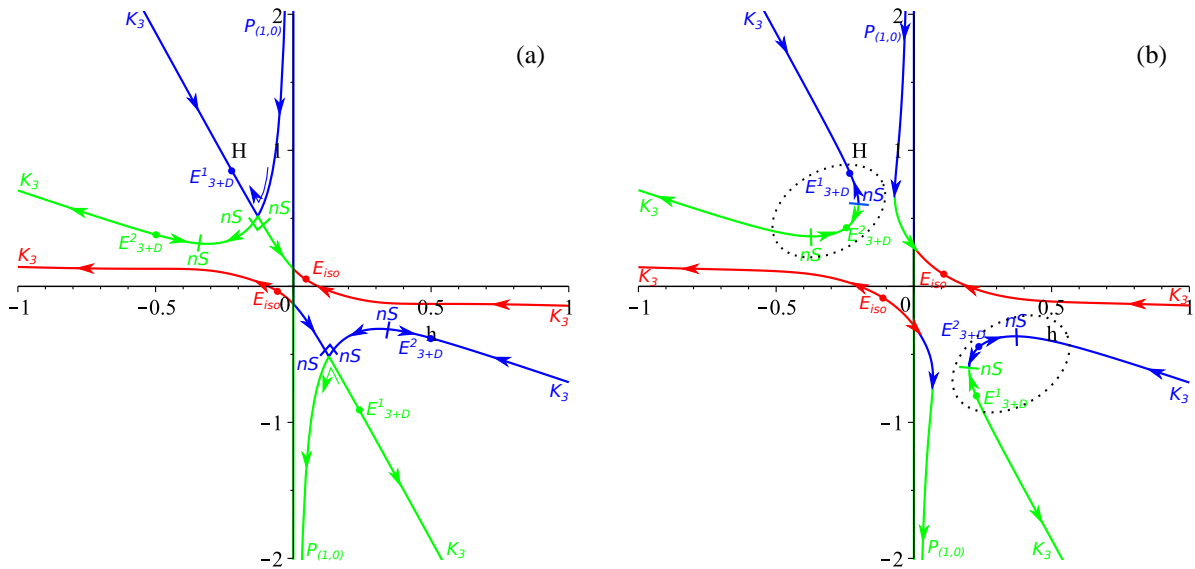}
\caption{({\bf{a}})--({\bf{b}}) Additional regimes shown in the resulting $H(h)$ graph for the $D\geqslant 4$ $\Lambda$-term case
(see the text for more details).}\label{D4Ladd} 
\end{figure}

Finally we have the general $D \geqslant 4$ case. The~results are presented in Figures~\ref{D4L} and~\ref{D4Ladd}. The~panel layout for Figure~\ref{D4L} is as follows: panel (a) corresponds to $(\alpha < 0, \Lambda < 0)$,
panel (b) to $(\alpha < 0, \Lambda > 0, \alpha\Lambda < \xi_3)$, panel (c) to $(\alpha < 0, \Lambda > 0, \xi_1 > \alpha\Lambda > \xi_3)$, panel (d) to $(\alpha < 0, \Lambda > 0, 0 > \alpha\Lambda > \xi_1)$,
panel (e)~ to $(\alpha > 0, \Lambda < 0)$ and panel (f) to $(\alpha > 0, \Lambda > 0, \alpha\Lambda < \xi_2)$. The~panels in Figure~\ref{D4Ladd} are as follows: panel (a) is for $(\alpha > 0, \Lambda > 0, \alpha\Lambda = \xi_2)$ while panel
(b) is for $(\alpha > 0, \Lambda > 0, \alpha\Lambda > \xi_2)$.

The quoted $\xi_{1, 2, 3}$ are exact values found in~\cite{my17a} (where they are denoted as $\zeta_{1, 2, 3}$):

\begin{myequation}
\begin{array}{l}
\xi_1 = - \dac{(D+2)(D+3)}{4D(D+1)},~\xi_2 = \dac{\sqrt[3]{\mathcal{D}_2 (D-1)^2}}{12(D-2)(D-1)D(D+1)} + \\ \\ + \dac{(D^6 - 6D^5 + 10D^4 - 20D^2 + 24D + 36)(D-1)}{3D(D-2)(D+1)\sqrt[3]{\mathcal{D}_2 (D-1)^2}} +
\dac{D^3 - 9D^2 + 8D + 24}{12D(D-2)(D+1)},~\mbox{where} \\ \\
\mathcal{D}_2 = 10D^{10} + 6D^9 \mathcal{D}_1 - 100D^9 - 30D^8 \mathcal{D}_1 + 330 D^8 + 30D^7 \mathcal{D}_1 - 240 D^7 + 54D^6 \mathcal{D}_1 - \\ - 600D^6 - 84D^5 \mathcal{D}_1 + 240D^5 - 24D^4 \mathcal{D}_1 + 1520D^4 +
48D^3 \mathcal{D}_1 + 640 D^3 - 2880D^2 + 1728~\mbox{and}~ \\ \\ \mathcal{D}_1 = \dac{(D-4)(D-3)(D+2)}{(D-1)(D+1)}\sqrt{\dac{(D-4)(D+2)}{D(D-2)}}.
\end{array} \label{D.4_sep1}
\end{myequation}

\begin{equation}
\begin{array}{l}
\xi_3 = - \dac{D(D-1)}{4(D-2)(D-3)}.
\end{array} \label{D.4_sep2}
\end{equation}

Let us have a look at the resulting regimes and find realistic ones among them; remember that $(H > 0, h < 0)$ regimes lie in the second quadrant. We have $K_3 \to K_3$ from Figure~\ref{D4L}a $(\alpha < 0, \Lambda < 0)$ as a
potentially realistic regime. Further, there is $K_3 \to E^1_{3+D}$ and $P_{(1, 0)} \to E^1_{3+D}$ from Figure~\ref{D4L}b $(\alpha < 0, \Lambda > 0, \alpha\Lambda \leqslant \xi_1)$. This is the
``replacement'' of the discussed situation ``moving'' the exponential solution for $D=2, 3$, which ``hits'' zero at $\alpha\Lambda = -3/2$, but in the general $D \geqslant 4$ case it hits at $\alpha\Lambda = \xi_1$. The~same
realistic regimes---$K_3 \to E^1_{3+D}$ and $P_{(1, 0)} \to E^1_{3+D}$---can be found in Figures~\ref{D4L}e $(\alpha > 0, \Lambda < 0)$, \ref{D4L}f $(\alpha > 0, \Lambda > 0, \alpha\Lambda < \xi_2)$ and \ref{D4Ladd}a
$(\alpha > 0, \Lambda > 0, \alpha\Lambda = \xi_2)$. From Figure~\ref{D4Ladd}b we can see that only $K_3 \to E^1_{3+D}$ can be called realistic in the $(\alpha > 0, \Lambda > 0, \alpha\Lambda > \xi_2)$ case,  but, similar to the
previous cases, it has the fine structure of anisotropic exponential solutions (see~\cite{my17a} for details) and $K_3 \to E^1_{3+D}$ exists only for $\alpha\Lambda  \leqslant \xi_4$ where $\xi_4$ is found
in~\cite{my17a} (where it is denoted as $\zeta_{6}$):

\begin{equation}
\begin{array}{l}
\xi_4 = \dac{1}{4} \dac{3D^2 - 7D + 6}{D(D-1)}.
\end{array} \label{D.4_sep3}
\end{equation}

To conclude, the~general $D \geqslant 4$ case is quite similar to $D=3$ but the limits are different. We can call the following regimes realistic: $K_3 \to K_3$ for ($(\alpha < 0, \Lambda < 0)$); and
$K_3 \to E_{3+D}$ and $P_{(1, 0)} \to E_{3+D}$ for $(\alpha < 0, \Lambda > 0, \alpha\Lambda \leqslant \xi_1)$ and $(\alpha > 0, \alpha\Lambda \leqslant \xi_2)$ (including the entire $\Lambda < 0$ domain).
 For $K_3 \to E_{3+D}$
alone the second region is extended to $\alpha\Lambda  \leqslant \xi_4$.

Overall, the~general $D \geqslant 4$ $\Lambda$-term case demonstrates the richest dynamics of all cases considered so far. It also has the widest areas on the parameters space, which could lead to realistic compactification regimes.

\subsection{Conclusions on the $\Lambda$-Term Case}

To conclude, the~richness of the dynamics in different $D$ cases increases with the growth of $D$, as in the vacuum case. Unlike the vacuum case, where we report almost no changes for all $D \geqslant 2$ cases, the~$\Lambda$-term
cases have steady growth of the regime abundance with increasing $D$. In~$D=1$ we have only one would-be realistic regime---$P_{(1, 0)}\to P_{(0, 1)}$, which exists for $(\alpha > 0, \Lambda < 0)$. The~situation with
$D=2$ changes drastically. This regime is no more but there are two others, $K_3 \to E_{3+D}$ and $P_{(1, 0)} \to E_{3+D}$. The former exists for $(\alpha > 0, \alpha\Lambda \leqslant 1/2)$
(including the entire $\Lambda < 0$ domain) while the latter exists for $(\alpha > 0, \alpha\Lambda < -1/6)$. In~$D = 3$ we have the additional regime $K_3 \to K_3$ which exists for $(\alpha < 0, \Lambda < 0)$ and two regimes which are
the same as in the $D=2$ case, $K_3 \to E_{3+D}$ and $P_{(1, 0)} \to E_{3+D}$. They both exist for $(\alpha < 0, \Lambda > 0, \alpha\Lambda \leqslant -3/2)$ and $(\alpha > 0, \Lambda < 0)$; for $K_3 \to E_{3+D}$ the
area of existence is extended to $(\alpha > 0, \Lambda > 0, \alpha\Lambda \leqslant 1/2)$. Finally for the general $D \geqslant 4$ case we have all the regimes from $D=3$ but with partially different areas of existence:
$K_3 \to K_3$ has the same area of existence $(\alpha < 0, \Lambda < 0)$ while $K_3 \to E_{3+D}$ and $P_{(1, 0)} \to E_{3+D}$ exist for $(\alpha < 0, \Lambda > 0, \alpha\Lambda \leqslant \xi_1)$ and
$(\alpha > 0, \alpha\Lambda \leqslant \xi_2)$ (including the entire $\Lambda < 0$ domain). For $K_3 \to E_{3+D}$ alone the second region is extended to $\alpha\Lambda  \leqslant \xi_4$.

\section{Discussion}

Let us start the discussion of the obtained results with the discussion of the asymptotes. For the past, there are two of them---$K_3$ and $P_{(1, 0)}$. Both of them are power law ($a(t) \propto t^{p}$) regimes, and~the difference
lies in the expansion rates---for $K_3$ we have $p_H$ and $p_h$ so that $\sum p_i = 3p_H + D p_h = 3$ and $\{p_H, p_h\}\ne 0$ (here $p_H$ stands for the Kasner exponent associated with $H$ while $p_h$---for $h$).
For $P_{(1, 0)}$ it is the opposite situation---for this we have $p_H = 1$ and $p_h = 0$. Let us note that
formally for $P_{(1, 0)}$ $\sum p_i = 3$, and that is why we confused it with $K_3$ in the original papers~\cite{my16a, my16b, my17a}; in this paper we clearly separate them. The difference between them is in the
expansion rates. For $K_3$ we usually have $H > 0$ and $h < 0$, while for $P_{(1, 0)}$ it is more ``singular-like''---$H\to\pm\infty$ while $h\to 0\pm 0$.
We also note that $P_{(1, 0)}$ could be treated as a generalized Taub~\cite{Taub} solution. The~same is true for $P_{(0, 1)}$, which differs in the swapping of ``0'' and ``1'' in the Kasner exponents---for $P_{(0, 1)}$ we have
$p_H = 0$ (and so asymptotically $H\to 0\pm 0$) and $p_h = 1$ (and so $h\to\pm\infty$). A standard Kasner solution with $\sum p_i = \sum p_i^2 = 1$ finalizes the power law regimes presented in the course of study.

There is one important comment on the power law regimes as future asymptotes. In~\cite{my16a, my16b, my17a} we noted that $K_3$, being the future asymptote, is singular and has finite-time future singularity. The~same is
true for $P_{(1, 0)}$ as a future asymptote, at least in the $D=1$ $\Lambda$-term case (see~\cite{my16b}). The~reason behind it not clear and requires additional investigation, but the fact is stated, and~so we remove
the regimes with $K_3$ and $P_{(1, 0)}$ as future asymptotes from the list of realistic regimes.

With little to comment on with respect to exponential regimes, we turn our attention to another regime which is alien to GR---nonstandard singularity. It can be described as follows: in GR, being linear in the highest derivative,
if we solve the dynamical equations with respect to $\dot H_i$, the~resulting expressions are polynomial. On the contrary, for GB and other non-linear theories, the~resulting expression is rational function. There could
be a situation when the denominator of this function hits zero at some regular $H$ while the numerator is regular and nonzero. In~this case the result diverges and so $\dot H_i$, making this point singular. However, it happens
at regular nonzero $H$, making the situation nonstandard and providing the origin of the name for singularity of this type. Let us note that this kind of singularity is quite common in GB. For the totally anisotropic
spatially flat vacuum $(4+1)$-dimensional model it is the only future asymptote (see~\cite{prd10}). This kind of singularity is ``weak'' as per Tipler's classification~\cite{Tipler}, and~``type II''
in classification as per Kitaura and Wheeler~\cite{KW1, KW2}.

\begin{figure}
\includegraphics[width=1.0\textwidth, angle=0]{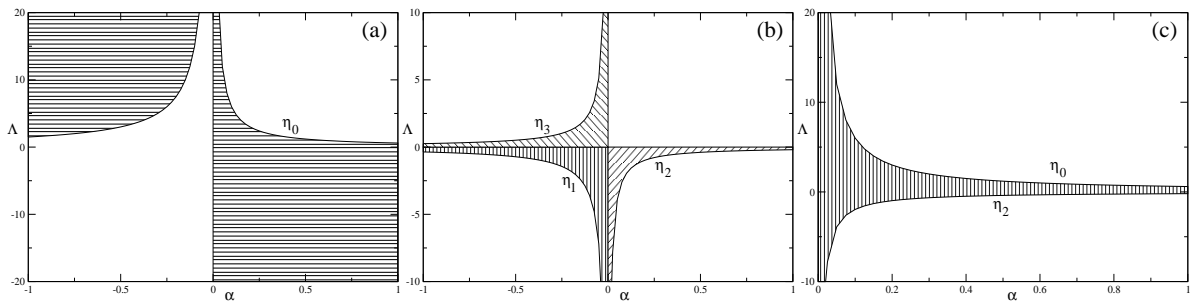}
\caption{Summary of the bounds on ($\alpha, \Lambda$) from this paper alone in panel (\textbf{a}); from other considerations found in the literature in panel (\textbf{b}); and the intersection between them
in panel (\textbf{c})
(see the text for more details).}\label{DD}
\end{figure}

We have found domains on the ($\alpha$, $\Lambda$) plane where realistic regimes exist and it is interesting to compare the bounds  we have found with those coming from other considerations. The~results
of this comparison are presented in Figure~\ref{DD}. In~Figure \ref{DD}a we presented the summary of the results from the current paper---$\alpha < 0$, $\Lambda > 0$,
$\alpha\Lambda \leqslant \xi_1$ from (\ref{D.4_sep1}) in the second quadrant, and $\alpha > 0$, $\alpha\Lambda \leqslant \eta_0 \equiv \xi_4$ from  (\ref{D.4_sep3}) on the $\alpha > 0$ half-plane. In~Figure \ref{DD}b
we collected available constraints on $\alpha\Lambda$ from other considerations. Among them a significant number are based on the different aspects of Gauss--Bonnet gravity in AdS spaces---from the consideration of the shear viscosity to entropy ratio as well as causality violations and CFTs
 in the dual gravity description, limits were obtained
on $\alpha\Lambda$~\cite{alpha_01, alpha_02, alpha_03, alpha_04, alpha_05, alpha_06, alpha_07, alpha_08}:
\begin{equation}
\begin{array}{l}
 - \dac{(D+2)(D+3)(D^2 + 5D + 12)}{8(D^2 + 3D + 6)^2} \equiv \eta_2 \leqslant \alpha\Lambda \leqslant \eta_1 \equiv \dac{(D+2)(D+3)(3D + 11)}{8D(D+5)^2}.
\end{array} \label{alpha_limit}
\end{equation}

The limits for dS ($\Lambda > 0$) are less numerous and are based on different aspects (causality violations, perturbation propagation, and so on) of black hole physics in dS spaces. The~most
stringent constraint coming from these considerations is~\cite{add_rec_2, add_rec_4, dS}

\begin{equation}
\begin{array}{l}
\alpha\Lambda \geqslant \eta_3 \equiv - \dac{D^2 + 7D + 4}{8(D-1)(D+2)}.
\end{array} \label{alpha_limit2}
\end{equation}

At this point, two clarifications are required. First, (\ref{alpha_limit2}) is true for both $\alpha\lessgtr 0$ and $\Lambda \lessgtr 0$, so that in the $\alpha > 0$, $\Lambda < 0$ quadrant two limits
are applied: $\alpha\Lambda \geqslant \eta_2$ from (\ref{alpha_limit}) and $\alpha\Lambda \geqslant \eta_3$ from (\ref{alpha_limit2}). One can easily check that $\eta_2 > \eta_3$ for
$D \geqslant 2$ so that the constraint from (\ref{alpha_limit}) is the most stringent in this quadrant. Secondly, one can see that the limit in (\ref{alpha_limit2}) is not defined for
$D=1$. Indeed, in~this case the limit is special (see~\cite{alpha_12}), but since for $D=1$ there are no viable cosmological regimes, we consider $D \geqslant 2$ only.

Secondly, one can see that the bounds on ($\alpha, \Lambda$) cover three quadrants and our analysis allows for constraint of the remaining sector: $\alpha > 0$, $\Lambda > 0$. However, if we consider joint
constraint from Figure~\ref{DD}a,b, the~resulting area is presented in Figure~\ref{DD}c. In~there, one can see that the regimes in the $\alpha < 0$ sector disappear due to the fact that
$\eta_3 > \xi_1$ always---the ($\alpha, \Lambda$) values, which have viable cosmological dynamics in $\alpha < 0$ sector, disagree with (\ref{alpha_limit2}). To conclude, if we consider our bounds on
($\alpha$, $\Lambda$) together with previously obtained (see (\ref{alpha_limit})--(\ref{alpha_limit2})), the~resulting bounds are
\begin{equation}
\begin{array}{l}
\alpha > 0, \quad D \geqslant 2, \quad \dac{3D^2 - 7D + 6}{4D(D-1)}  \equiv \eta_0 \geqslant \alpha\Lambda \geqslant \eta_2 \equiv - \dac{(D+2)(D+3)(D^2 + 5D + 12)}{8(D^2 + 3D + 6)^2}.
\end{array} \label{alpha_limit3}
\end{equation}

The result that the joint analysis suggests only $\alpha > 0$, which is interesting and important--- indeed, the~constraints on ($\alpha, \Lambda$) considered so far do not distinguish between
$\alpha \lessgtr 0$, and~there are several considerations which favor $\alpha > 0$. The~most important of them is the positivity of $\alpha$ coming from heterotic string setup, where $\alpha$
is associated with inverse string tension~\cite{alpha_12}, but there are several others like ill-definition of the holographic entanglement entropy~\cite{entr}. Hence, our joint analysis
supports $\alpha > 0$ as well.

\section{Conclusions}

In this paper we provide a comprehensive review of the situation of realistic compactification in the spatially flat Einstein--Gauss--Bonnet cosmologies. It is not just a repetition of~\cite{my16a, my16b, my17a} but a
brand new approach to the problem. The~original papers contain a lot of technical details---they are necessary, as they provide proof of the solidness of the results. However, the abundance of the technical details sometimes hides the
meaning, and in the present paper we unveil it. We introduce a new presentation of the resulting regimes---placing them all on $H(h)$ or $h(H)$ curves and indicating the direction of the evolution with arrows. Finally, we
corrected ourselves with respect to the wrong treatment of the $P_{(1, 0)}$ regime. Our analysis allows us to put constraints on the previously unconstrained sector on the ($\alpha, \Lambda$) plane---$\alpha > 0$, $\Lambda > 0$. Joint analysis
of our bounds on the ($\alpha, \Lambda$) with those obtained from other considerations allow us to tighten the constraints on $\alpha\Lambda$ (see Equation~(\ref{alpha_limit3})) and drop $\alpha < 0$ from consideration.
However, the field still has a lot of unresolved mysteries, like the mentioned singular behavior of $K_3$ and $P_{(1, 0)}$ when they are future asymptotes. We will address this issue in due time.

\conflictsofinterest{The authors declare no conflict of interest.}




\end{document}